\documentclass[11pt]{article}
\usepackage{moriond,epsfig}

\def\Journal#1#2#3#4{{#1} {\bf #2}, #3 (#4)}

\def\NPB{{\em Nucl. Phys.} B}
\def\PLB{{\em Phys. Lett.}  B}
\def\PRL{\em Phys. Rev. Lett.}

\def\be{\begin{equation}}
\def\ee{\end{equation}}
\def\bea{\begin{eqnarray}}
\def\eea{\end{eqnarray}}

\def\ra{\rightarrow}

\def\dkln{D^0 \rightarrow K^-l^+\nu_l}
\def\dkpln{D^+ \rightarrow K^-\pi^+l^+\nu_l}

\def\dkskpln{D^+ \rightarrow \overline{K}^{*0}(K^-\pi^+)l^+\nu_l}
\def\dsphikkln{D_s^+ \rightarrow \phi(K^-K^+)l^+\nu_l}

\def\dkpmn{D^+ \rightarrow K^-\pi^+\mu^+\nu}

\def\dkskpmn{D^+ \rightarrow \overline{K}^{*0}(K^-\pi^+)\mu^+\nu}
\def\dsphikkmn{D_s^+ \rightarrow \phi(K^-K^+)\mu^+\nu}
\def\costv{\cos \theta_V}
\def\dksmn{D^+ \rightarrow \overline{K}^{*0}\mu^+\nu}
\def\dkpp{D^+ \rightarrow K^-\pi^+\pi^+}

\def\dsphikkp{D_s^+ \rightarrow \phi(K^-K^+)\pi^+}
\def\knp{K\,\pi}

\begin{document}
\vspace*{4cm}
\title{NEW RESULTS FROM THE FOCUS/E831 EXPERIMENT}

\author{ DORIS YANGSOO KIM \\
         Representing the FOCUS Collaboration
        \footnote{Collaboration members are
    J.~M.~Link, P.~M.~Yager (UC Davis);
    J.~C.~Anjos, I.~Bediaga C.~G\"obel, J.~Magnin, A.~Massafferri, 
    J.~M.~de~Miranda, I.~M.~Pepe, E.~Polycarpo, A.~C.~dos~Reis (CBPF);
    S.~Carrillo, E.~Casimiro, E.~Cuautle, A.~S\'anchez-Hern\'andez,
    C.~Uribe, F.~V\'azquez (CINVESTAV);
    L.~Agostino, L.~Cinquini, J.~P.~Cumalat, B.~O'Reilly, I.~Segoni,
    M.~Wahl (Colorado Boulder);
    J.~N.~Butler, H.~W.~K.~Cheung, G.~Chiodini, I.~Gaines,
    P.~H.~Garbincius, L.~A.~Garren, E.~Gottschalk, P.~H.~Kasper,
    A.~E.~Kreymer, R.~Kutschke, M.~Wang (FNAL); 
    L.~Benussi, L.~Bertani, S.~Bianco, F.~L.~Fabbri,
    A.~Zallo (INFN Frascati);
    M.~Reyes (Guanajuato); 
    C.~Cawlfield, D.~Y.~Kim, A.~Rahimi, J.~Wiss (Illinois Urbana-Champaign);
    R.~Gardner, A.~Kryemadhi (Indiana Bloomington); 
    Y.~S.~Chung, J.~S.~Kang, B.~R.~Ko, J.~W.~Kwak, 
    K.~B.~Lee (Korea University);
    K.~Cho, H.~Park (Kyungpook);
    G.~Alimonti, S.~Barberis, M.~Boschini, A.~Cerutti,  P.~D'Angelo,
    M.~DiCorato, P.~Dini, L.~Edera, S.~Erba, M.~Giammarchi,
    P.~Inzani, F.~Leveraro, S.~Malvezzi, D.~Menasce, M.~Mezzadri,
    L.~Moroni, D.~Pedrini, C.~Pontoglio, F.~Prelz, M.~Rovere,
    S.~Sala (Milano and INFN Milano);
    T.~F.~Davenport~III (North Carolina  Asheville);
    V.~Arena, G.~Boca, G.~Bonomi, G.~Gianini, G.~Liguori, D.~Lopes~Pegna,
    M.~M.~Merlo, D.~Pantea, S.~P.~Ratti, C.~Riccardi, P.~Vitulo 
     (Pavia and INFN Pavia);
    H.~Hernandez, A.~M.~Lopez, E.~Luiggi, H.~Mendez, A.~Paris, J.~Quinones,
    J.~E.~Ramirez, Y.~Zhang (Puerto Rico  Mayaguez);
    J.~R.~Wilson (South Carolina Columbia);
    T.~Handler, R.~Mitchell (Tennessee Knoxville);
    D.~Engh, M.~Hosack, W.~E.~Johns, M.~Nehring, P.~D.~Sheldon, K.~Stenson,
    E.~W.~Vaandering, M.~Webster (Vanderbilt); 
    M.~Sheaff (Wisconsin Madison)}
}

\address{Loomis Laboratory of Physics, 1110 W. Green St.\\
Urbana, Illinois, 61801, USA}

\maketitle\abstracts{
The E831/FOCUS experiment at Fermilab is a photoproduction experiment
which generated high quality charm particles. During its run, we
obtained a large data set, including more than 1 million charm mesons 
in the $K\pi/K2\pi/K3\pi$ mode decays. The current analysis efforts by the
collaboration members are quite active and diverse. 
I will summarize the recent papers published by the FOCUS group on 
topics of semileptonic decays of charm mesons.}

\section{Introduction}

In this paper, we will summarize three recent 
papers~\cite{focus_sw,focus_br,focus_ff} published 
by the FOCUS/E831 collaboration on topics of semileptonic decays of
charm mesons.  
 
\subsection{Semileptonic Decays of Charm Particles} 
Traditionally, the semileptonic decays of heavy flavored particles are
accessible to both collider and fixed target experiments with ease. 
The decays have clean and distinguishable signatures, and the Cabbibo-allowed
decay channels like $\dkln$, $\dkskpln$, $\dsphikkln$ 
have large branching ratios. 

Their fully explicit decay rates can be calculated from first principles,
for example, theoretical tools like Feynman diagrams. Involving a lepton
in the final decay stage implies that we do not have to worry about the usual
final state interaction between hadrons.  The possible complications coming
from the QCD portion of the decay process are contained in form factors.
The form factors can be calculated by various methods, LGT and quark
models. The angular distributions and invariant masses among the decay
products would determine the form factors ratios while the branching
ratio measurements and information from the CKM matrix would give the absolute
scale for the form factors.  

\subsection{The FOCUS Spectrometer}
The FOCUS/E831 spectrometer is an upgraded version of the E687 fixed target
spectrometer located in the Fermilab proton beam area. It is designed to study
charm particles produced by average 180 $GeV$ photon beams and BeO target
segments. The spectrometer consists of a precise silicon micro vertex system,
proportional wire chambers with two dipole magnets, threshold Cerenkov systems,
electromagnetic and hadronic calorimeters, and muon identification systems.
During the 1996-1997 fixed target run, a huge data set was obtained including
more than 1 million charm mesons in the $K\pi/K2\pi/K3\pi$ decay modes.
The members of the E831 group are from diverse countries, USA (and Puerto Rico),
Italy, Brazil, Mexico and Korea.

\section{The New S-wave Interference in $\dkpmn$ Decays}
For last 20 years, people regarded the $\dkpmn$ decays as 100\% $\dkskpmn$
events. The E687 and E691 groups set an upper limit for the possible scalar 
contributions in the  $\dkpln$ decays~\cite{e687,e691}, 
but they could not provide clear evidence of decay paths
other than the dominant P-wave $\dkskpln$ channel.
The situation was changed when the next generation sample from the FOCUS
spectrometer was analyzed to get form factors of the 
$\dkpmn$ decays~\cite{focus_sw}.

After the selection cuts involving vertex confidence levels and particle 
identification requirements, we obtained 31,254 $\dkpmn$ and its 
charge conjugate decays\footnote{In this paper we assume that a decay and
its charge conjugate decay go through the same physical process.}. 
During the form factor analysis,
we checked the angular distribution of Kaon in the $\knp$ rest frame
($\costv)$ and found that it showed a huge forward-backward asymmetry
below the $K^*(892)$ pole mass while almost no asymmetry
above the pole.  Since the $K^*$ is a P-wave, pure
$K^* \ra K\pi$ decays would have shown only a symmetric forward-backward
$\costv$ distribution over the entire $\knp$ invariant mass range. This 
suggests a possible quantum mechanics interference effect.  

   A simple approach to emulate the interference effect is adding a
spin zero amplitude in the matrix elements of the $\dkpmn$ decays. We tried 
a constant amplitude with a phase, $A \exp(i\delta)$, in the place
where the $K^*$ couples to the spin zero component of the $W^+$ particle. 
We made the simplest assumption that the $q^2$ dependence of this anomaly S-wave
coupling would be the same as that of the $K^*$.

\begin{figure}

\begin{center}
\psfig{figure=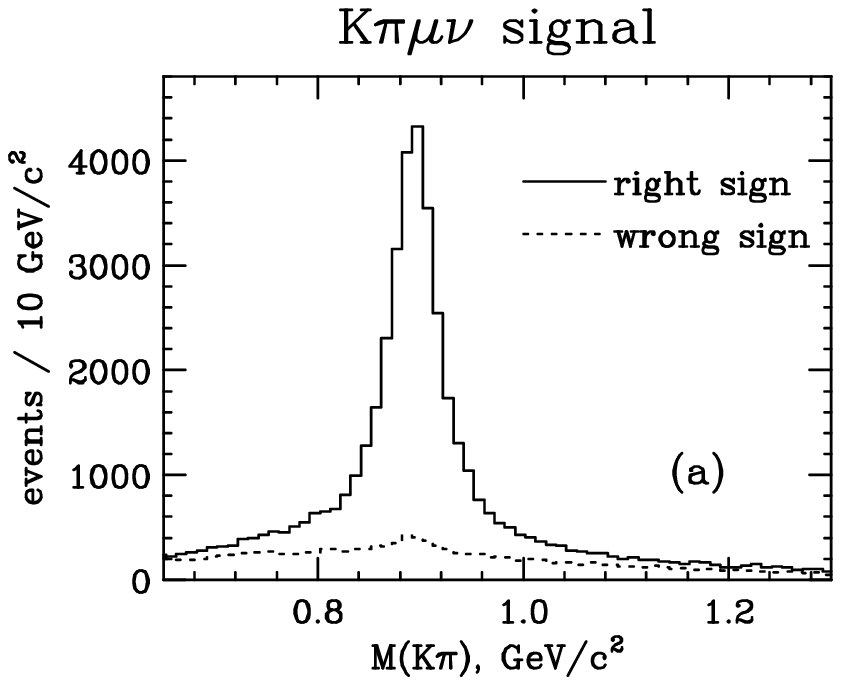,height=2.0in} 
\hspace{0.5in}
\psfig{figure=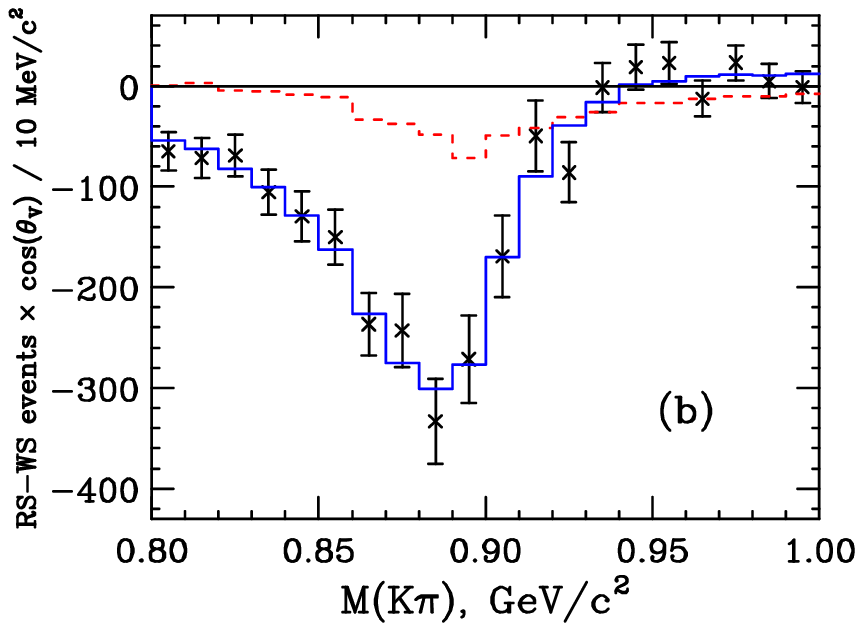,height=2.0in} 
\end{center}

\caption{(a) $\dkpmn$ signal. The wrong-sign-subtracted yield
is 31,254 events. (b) Asymmetry distribution in $\knp$ invariant
mass. The dashed line 
represents Monte Carlo simulation with no interfering s-wave amplitude
while the solid line represents Monte Carlo simulation with an s-wave
amplitude. The points with error bars are the experimental data. The figures
and the captions are reproduced from Reference~1.
\label{fig:costv_mkpi}}

\end{figure}

   The $\dkpmn$ event is a 4-body decay, which is represented by 5 kinematic
variables, two invariant masses and three angular variables. For each of
these variables, we extracted interference effects by using various weighting
schemes and studied if our measured $A = 0.36$ and $\delta = \pi/4$
are working properly in reproducing the effects for Monte Carlo (MC) 
events~\cite{focus_sw}. 
As shown in Fig.~\ref{fig:costv_mkpi}
where the invariant mass of the $\knp$ particles are weighted by $\costv$,
the interference effect is reproduced with satisfaction. Our measured phase
of $\pi/4$ relative to the $K^*(892)$ is consistent with the one found by LASS
collaboration for isosinglet s-wave around the K* pole from a $\knp$ phase
shift analysis~\cite{lass}. Our data is consistent with a broad 
resonance interpretation as well, but the pole of the resonance would be located
above the $K^*$ pole in absence of any FSI re-phasing.
We tried a $\kappa(800)$ resonance hypothesis. It turned out that 
to produce the interference effect, a 100 degree phase shift is needed
between the $\kappa$ and the $K^*$.

One interesting side effect of the S-wave interference is that it breaks 
the $\chi \leftrightarrow -\chi$ symmetry of the distribution
of the azimuthal angle ($\chi$)
between the $\knp$ and the $W^+$ decay planes in the $D^+$ rest frame. 
The proper definition of $\chi$ requires that it should change sign
between $\dkpmn$ and its charge conjugate decays. Without the proper sign
convention, we would see a false CP violation between the charge conjugate
decays in the $\chi$ distribution.

\section{Branching Ratio Measurements}
    We measured the relative branching ratio between $\dksmn$ and $\dkpp$
decays.  With a tighter selection than the one used in the interference
analysis, we selected 11,698 $\dkpmn$ and its charge conjugate decays. With a
selection cut set designed to be similar to the one applied upon the
$\dkpmn$ decays, we obtained 65,421 $\dkpp$ and its charge conjugate decays.
From a MC study, we determined that the pure $\dksmn$ events are 94.5\%
of the selected events. When this correction factor is applied, we
obtained~\cite{focus_br},
\be
   \frac{\Gamma(\dksmn)}{\Gamma(\dkpp)} =
      0.602 \pm 0.010 (stat) \pm 0.021 (sys) 
\ee
When comparing this muon decay channel result with electron decay channel
results from other experiments, a correction factor 1.05 should be applied. Our
number, the only one considered an S-wave interference explicitly, is
1.6 $\sigma$ lower than the recent CLEO II result 
from the electronic decay channel~\cite{cleo} and 2.1 $\sigma$ higher than 
the E691 measurement~\cite{e691_br}.  
Including our result, the new world average of
$\Gamma(K^*l\nu)/\Gamma(K\pi\pi)$ is 0.62 $\pm$ 0.02 each experiment's
statistical and systematic errors were added in quadrature
prior to making the weighted average.

We also measured the relative branching ratio between $\dsphikkmn$
and $\dsphikkp$ decays. Our selection yields 793 $\dsphikkmn$ and its charge
conjugate decays, and 2,192 $\dsphikkp$ and its charge conjugate decays.
The result is~\cite{focus_br} 
\be
   \frac{\Gamma(\dsphikkmn)}{\Gamma(\dsphikkp)} =
      0.540 \pm 0.033 (stat) \pm 0.048 (sys)
\ee
Our number is comparable with all the other measurements in this channel,
and the new world average of $\Gamma(\phi\mu\nu)/\Gamma(\phi\pi)$ is
0.540 $\pm$ 0.040. 

\section{The Form Factor Ratios of $\dksmn$}
We measured the form factor ratios of $\dksmn$ and it charge conjugate decays
with consideration on the S-wave contribution. Our study shows that the
effect of S-wave on the measurement is minimal while the effect of charm
background is significant. The new FOCUS results are as follows~\cite{focus_ff},
\bea
   & R_V = 1.504 \pm 0.057 \pm 0.039 \\
   & R_2 = 0.875 \pm 0.049 \pm 0.064
\eea 
Our $R_V$ value is 2.9 $\sigma$ below the E791 measurements~\cite{e791},
but consistent with others. Our $R_2$ value is consistent with other
measurements. The new world averages are 1.66 $\pm$ 0.060 and
0.827 $\pm$ 0.055 for $R_V$ and $R_2$, respectively. 

\section{Summary and Future Plan}
The FOCUS experiment found new S-wave interference phenomena in the
$\dkpmn$ decays. Considering this effect in further analyses, we measured
the branching ratio $\Gamma(D^+ \ra K^*\mu\nu)/\Gamma(D^+ \ra K\pi\pi)$ 
and the form factor ratios of $\dkpmn$ decays with improved statistical errors. 
We also measured the branching ratio 
$\Gamma(D_s \ra \phi\mu\nu)/\Gamma(D_s \ra \phi\pi)$. 
The analyses in other semileptonic decay modes are actively going on
and we expect new results soon. 

\section*{Acknowledgments}
We wish to acknowledge the assistance of the staffs of Fermi National
Accelerator Laboratory, the INFN of Italy, and the physics departments of the
collaborating institutions. This research was supported in part by the U.~S.
National Science Foundation, the U.~S. Department of Energy, the Italian
Istituto Nazionale di Fisica Nucleare and Ministero della Istruzione
Universit\`a e Ricerca, the Brazilian Conselho Nacional de Desenvolvimento
Cient\'{\i}fico e Tecnol\'ogico, CONACyT-M\'exico, and the Korea Research
Foundation of the Korean Ministry of Education.

\end{document}